\documentclass[pdflatex,sn-mathphys-num]{sn-jnl}

\usepackage{graphicx}%
\usepackage{subcaption}
\usepackage{multirow}%
\usepackage{amsmath,amssymb,amsfonts}%
\usepackage{amsthm}%
\usepackage{mathrsfs}%
\usepackage[title]{appendix}%
\usepackage{xcolor}%
\usepackage{textcomp}%
\usepackage{manyfoot}%
\usepackage{booktabs}%
\usepackage{algorithm}%
\usepackage{algorithmicx}%
\usepackage{algpseudocode}%
\usepackage{listings}%
\usepackage{array}
\usepackage{caption,subcaption,tabularx,array}
\newcolumntype{L}[1]{>{\raggedright\arraybackslash}p{#1}} 

\theoremstyle{thmstyleone}%
\theoremstyle{thmstyletwo}%

\theoremstyle{thmstylethree}%

\begin{document}

\title[Article Title]{Identification of Authoritative Nodes and Dismantling of Illicit Networks Using a Novel Metric for Measuring Strength of a Graph}


\author{\fnm{Kartikeya} \sur{Kansal}}\email{kkansal1@asu.edu}
\author{\fnm{Arunabha} \sur{Sen}}\email{asen@asu.edu}





\affil{\orgdiv{School of Computing and Augmented Intelligence}, \orgname{Arizona State University}, \orgaddress{\street{699 S. Mill Avenue}, \city{Tempe}, \postcode{85048}, \state{Arizona}, \country{USA}}}









\abstract{
{\em Dismantling} criminal networks or containing epidemics or misinformation through node removal is a well-studied problem. To evaluate the {\em effectiveness} of such efforts, one must measure the {\em strength} of the network {\em before} and {\em after} node removal. Process $P_1$ is considered more effective than $P_2$ if the strength of the residual network after removing $k$ nodes via $P_1$ is {\em smaller} than that from $P_2$. This leads to the central question: How should network strength be measured?

Existing metrics rely solely on {\em structural} properties of the graph, such as {\em connectivity}. However, in real-world scenarios—particularly in law enforcement—the {\em perception} of agents regarding network strength can differ significantly from structural assessments. These perceptions are often ignored in traditional metrics.

We propose a new strength metric that integrates both structural properties and {\em human perception}. Using {\em human subject surveys}, we validate our approach against existing metrics. Our metric not only aligns more closely with human judgment but also outperforms traditional methods in identifying {\em authoritative nodes} and effectively dismantling both synthetic and real-world networks.
}




\maketitle

\section{Introduction}

Dismantling criminal networks, containing epidemics, or limiting the spread of misinformation by removing key nodes from a network has been a long-standing research challenge. Despite the diversity of application domains, the underlying problem remains consistent: {\em How can we identify the set of nodes whose removal most significantly weakens the network?} These nodes are often the most {\em authoritative/influential} actors. Seminal work by Kempe {\em et al.} addressed this in the context of influence maximization in social networks \cite{kempe2003maximizing}. More recently, De La Mora Tostado {\em et al.} investigated human trafficking networks in southern Mexico \cite{de2024modeling}, and Ren {\em et al.} proposed a generalized dismantling framework minimizing the cost of fragmentation \cite{ren2019generalized}.

Given the importance of this problem, a large body of literature has emerged \cite{kempe2003maximizing, de2024modeling, ren2019generalized, Shen2012, lu2016vital, patron2017optimal, braunstein2016network, albert2000error, vespignani2012modelling, piccini2016graph}. These works generally fall into two categories: {\em foundational} and {\em applied}. Foundational studies take either a Network Science approach \cite{ren2019generalized, lu2016vital, patron2017optimal, braunstein2016network, albert2000error, vespignani2012modelling} or a Graph-Theoretic/Combinatorial Optimization approach \cite{capobianco1979strength, chvatal1973tough, gusfield1991computing, trubin1993strength, Shen2012, piccini2016graph}. Irrespective of the methodology, assessing the {\em effectiveness} of any dismantling strategy requires comparing the {\em strength} of the network before and after node removal. For two processes $P_1$ and $P_2$, $P_1$ is more effective if the residual network after removing $k$ nodes is weaker than that resulting from $P_2$.

This raises a central question: {\em How do we quantify network strength or toughness?} Numerous metrics have been proposed \cite{MathWorldGraphStrength, WikipediaGraphToughness, capobianco1979strength, gusfield1991computing, chvatal1973tough, bauer2006toughness}, yet they are uniformly based on structural properties of the graph.

Our work is supported by an NSF grant investigating strategies to mitigate human trafficking in the U.S. Southwest. In collaboration with a large metropolitan police department, we observed a critical gap: the structural metrics used in literature do not always align with the perceptions of Law Enforcement Agents (LEAs). Field practitioners often judge network robustness differently than algorithms based purely on graph topology. This mismatch inspired us to integrate {\em human perception} directly into the definition of network strength.

In this paper, we propose a novel metric that combines structural features of the network with perceptual feedback collected from human subjects. Our metric introduces a {\em tunable weight vector} $W$ that reflects the perceived importance of connected component sizes. This vector is derived through surveys over a diverse set of synthetic and real-world networks. Our evaluation shows that the proposed metric correlates strongly with human intuition and surpasses traditional metrics in assessing network strength and identifying key nodes.

Criminal and terrorist networks are typically modeled as graphs, where nodes represent individuals and edges represent observed relationships (e.g., meetings, phone calls). LEAs often disrupt such networks by arresting key individuals and then assess the resulting network's degradation. If one disrupted network loses 25\% of its effectiveness, while another loses 50\%, the former is perceived as {\em tougher}. However, existing metrics do not account for how LEAs or domain experts interpret these differences. Our metric addresses this by aligning strength evaluation with human assessments.

\begin{figure*}[t!]
    \centering
    \includegraphics[width=5in,height=3in]{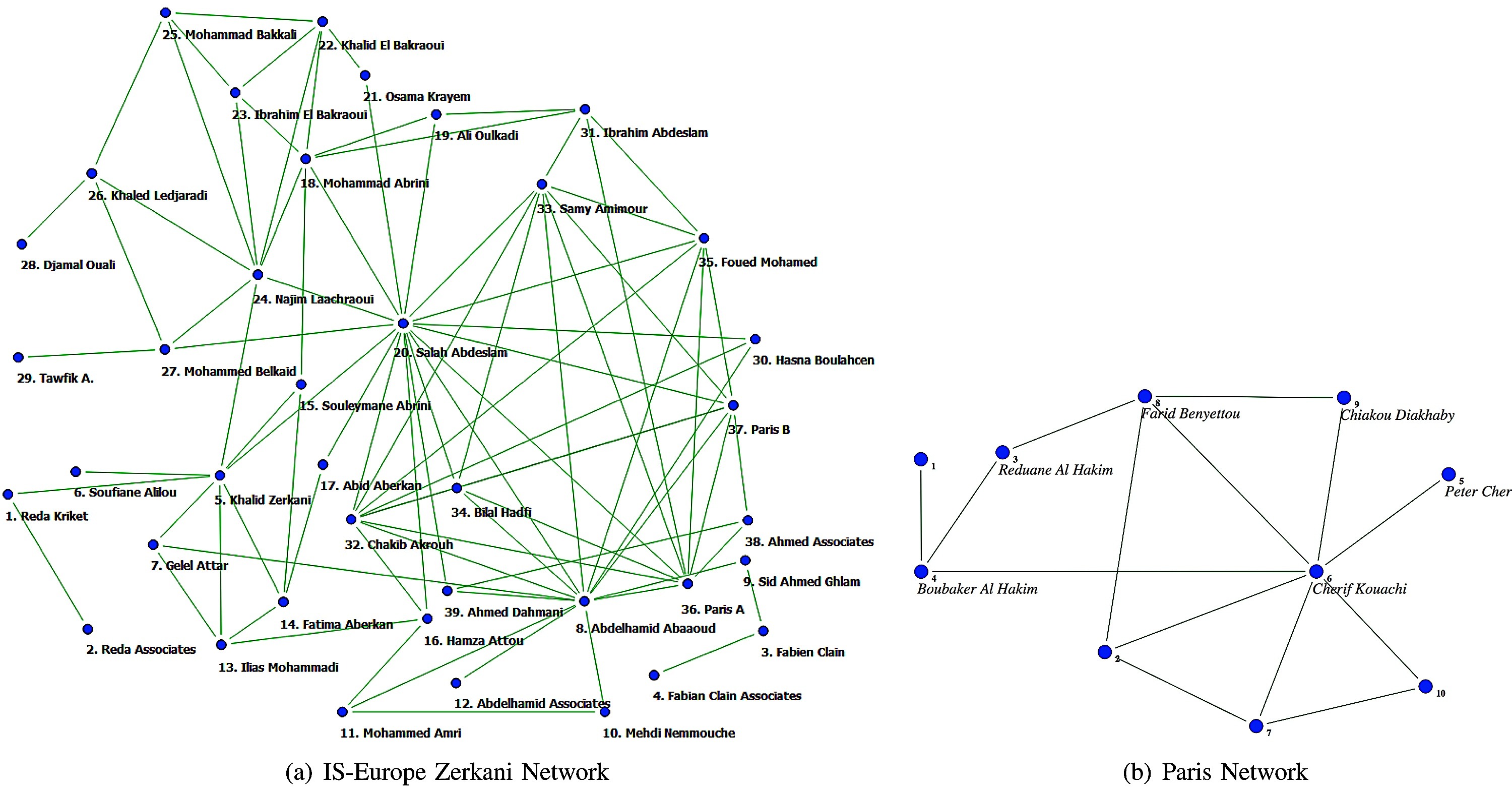}
    \caption{Terrorist network in Paris attack in 2015}
    \label{KaustavNetworks}
\end{figure*}

Figure \ref{KaustavNetworks} shows an example of a real-world network used in our study—illustrating the kind of illicit networks analyzed.

\vspace{2mm}
\noindent
\textbf{Our contributions are summarized below:}
\begin{itemize}
    \item We propose a new strength metric incorporating both structural properties and human perception.
    \item We use the metric to identify $k$ most authoritative nodes via an ILP-based optimization.
    \item We derive tunable weights based on human-subject responses across diverse networks.
    \item We evaluate our approach on synthetic and real criminal networks, demonstrating its superiority over existing metrics.
    \item We show that our metric better aligns with human perception than Cole 1, Cole 2 \cite{Shen2012}, and the GFP metric \cite{piccini2016graph}, which represent a class of metrics relying solely on structural properties of a network. In contrast, our metric integrates both structural information and human perception, leading to significantly improved performance.
\end{itemize}

\section{Related Work}

In a recent paper \cite{de2024modeling}, the authors studied dismantling human trafficking networks through graph-theoretic analysis. They constructed a network from Chiapas, Mexico using {\em snowball sampling}, resulting in 34 nodes and 225 edges. The nodes represented eleven actor types and edges denoted relationships. Since actors held varying importance, the network was modeled as a {\em node-weighted} undirected graph.

Dismantling such networks involves actions—e.g., arrests—that correspond to node removals. The goal is to reduce the network’s {\em strength}, making the development of appropriate metrics essential. Various definitions of network {\em strength} and {\em toughness} have emerged across disciplines \cite{MathWorldGraphStrength, WikipediaGraphToughness, capobianco1979strength, chvatal1973tough, gusfield1991computing, bauer2006toughness}.

Capobianco and Molluzzo \cite{capobianco1979strength} define strength via $\sigma_1 = 1/S.S$, where $S$ is a vector representing increases in connected component count upon node deletion. A more general formulation from Gusfield \cite{gusfield1991computing} is:
\[
\sigma_2(G) = \min_{S} \frac{|S|}{c(G - S) - 1},
\]
where $c(G - S)$ denotes the number of components after removing node set $S$. This metric reflects resistance to edge deletion.

Chvátal’s {\em toughness} \cite{chvatal1973tough} offers a related measure: a graph is $t$-tough if, for any integer $s > 1$, removing fewer than $t \cdot s$ vertices results in no more than $s$ components. This notion is particularly relevant when analyzing resilience to vertex removal.

Albert {\em et al.} \cite{albert2000error} showed that scale-free networks are highly tolerant to random failures but highly vulnerable to targeted attacks. Girvan and Newman \cite{girvan2002community} introduced {\em edge betweenness centrality} as a vulnerability indicator, while Holme {\em et al.} \cite{holme2002attack} examined attack effects on component size and path lengths.

Spectral methods, such as those explored by Chung \cite{chung1997spectral}, analyze eigenvalues of graph matrices (e.g., adjacency, Laplacian) to infer structural robustness. While powerful, these methods require expertise and are less intuitive for practitioners.

Simpler structural metrics are also prevalent: the size of the largest connected component and the number of components are commonly used as proxies for robustness. Shen {\em et al.} \cite{Shen2012} proposed methods to identify $k$ nodes whose removal minimizes the largest residual component and maximizes total fragmentation.

Piccini {\em et al.} \cite{piccini2016graph} introduced the Graph Fragmentation Problem (GFP), where the objective is to remove a subset $V^* \subseteq V$ within a budget $B$ to minimize:
\[
\mathrm{Score}(G') = \sum_{i=1}^k p_i n_i,
\]
where $p_i = n_i / n$ and $n_i$ is the size of each residual component. This score estimates the expected number of nodes affected if a failure begins at a randomly selected node.

While these metrics are analytically grounded, they rely solely on structural properties. In contrast, our proposed metric integrates human perception into the evaluation, offering a more practical and interpretable framework, especially in high-stakes contexts such as law enforcement.

\section{Proposed Novel Metric}

As indicated earlier, our goal in proposing a new metric for measurement of strength of a network is to avoid relying just on the structural properties of the network and to incorporate human perception of strength in it. In the following, we explain how we achieve our objective.

A graph $G = (V, E)$ may be {\em connected} or {\em disconnected}. The definition of the metric to measure strength of a network provided here is applicable to both connected and disconnected graphs. A graph in its most general version can be viewed as a collection (or a set) of {\em connected components}. If $|V| = n$, and the graph is composed of $k$ connected components, $C = [C_1, C_2, \ldots, C_k]$, then $\sum_{i = 1}^k |C_i| = n$, where $|C_i|$ denotes the {\em cardinality} of the component $C_i$. It may be noted that $1 \leq |C_i|  \leq  n$. Significant amount of results exist in the networking research literature that utilize the {\em Degree Distribution} (DD) of the graph to study various properties associated with it. 
Along the line of the {\em Degree Distribution} of a graph, in this paper, we introduce the notion of {\em Connected Component Size Distribution} (CCSD) of a graph $G = (V, E)$. CCSD is a vector of size $n$, where the $i$-th element of the vector denotes the {\em number of components of size} $i$, denoted by ${nc}_i$. Thus the CCSD vector is given by $[{nc}_1, {nc}_2, \ldots, {nc}_n]$, where $1 \leq {nc}_i \leq n$ is the number of components of size $i$. The CCSD vector for a connected graph with $n$ nodes is $[0, 0, \ldots, 1]$. The size (or the cardinality) of a component in a graph may have different significance (or importance) in different application environments. In general, a larger size component is usually viewed as being more ``important'' than a smaller size component. However, in our metric, not only we wanted to incorporate this property, we also wanted to incorporate human perception of strength. Accordingly, we associate a {\em weight} $w_i$ with a connected component of size $i$. This weight vector $W = [w_1, w_2, \ldots, w_n]$ is introduced in the metric to capture human perception of strength and the $w_i, 1 \leq i \leq n$ values are computed from the human feedback about network strength. Our procedure for computation of $w_i$ parameters is described in section \ref{sec:w_values}. Given a graph $G = (V, E)$ with its associated CCSD vector $C$ and a weight vector $W = \{w_1, w_2, \ldots, w_n\}$, we define the strength metric, $\sigma(G, W)$,  of the graph in the following way.

\[\sigma(G, W) = \sum_{i = 1}^n i \times w_i \times nC_i \] 
It may be noted that for a connected graph $G = (V, E)$, \(\sigma_3(G, W) = n \times w_n \), as $\forall i, 1 \leq i \leq n-1, nc_i = 0$ and $nc_n = 1$.\\

The CCSD vector, combined with a customizable weight vector  $W$, allows us to define a Strength Metric 
$(G, W)$ that can be adapted to different applications and use cases. This generalization enables the evaluation of network strength under various conditions and scenarios, providing a more flexible and comprehensive measure of network resilience and fragmentation. We use the proposed metric to measure (i) strength of networks and (ii) identification of $k$ most authoritative nodes in the network.

\section{Data Collection for Experiments} 

\subsection{Graphs for Experiments} \label{sec:rwgraphs}
To simulate a diverse set of network topologies for our analysis, we generated approximately 150 synthetic graphs using the \texttt{erdos\_renyi\_graph} and \texttt{gnm\_random\_graph} functions from the NetworkX Python library. These models allowed us to create both probabilistic and fixed-edge-count graph instances. Specifically, the number of nodes in these graphs varied from 3 to 50, capturing small to moderately sized networks. For the Erd\H{o}s--R\'enyi model, we experimented with a range of edge probabilities \( p \in [0.05, 0.5] \), thereby controlling the expected edge density and enabling the study of both sparse and denser connectivity regimes. In contrast, the \( G(n, m) \) model allowed us to precisely control the number of edges \( m \), providing a complementary mechanism to explore structural variability under fixed edge constraints.  Around 120 of these graphs were used to calculate the weight vector while the remaining were used to conduct a comparative analysis of the proposed metric and other traditional metrics against the ground truth.

We also used some real-world networks created by using the following datasets: Saxena Terror India \cite{Saxena2004}, Rhodes Bombing \cite{Rhodes2009}, Global Suicide Attacks \cite{Acosta2013}, Cocaine Dealing (Natarajan) \cite{Natarajan2006}, Cocaine Smuggling (ACERO) \cite{JimenezSalinas2011}, Human Trafficking Network (CHIAPAS) \cite{de2024modeling} and graphs (ZERKANI and PARIS) in Figure \ref{KaustavNetworks}.

\subsection{Weight Vector $W$ Computation} \label{sec:w_values}

In this section, we explain how we compute the weight vector $W$. This computation involves two parts. In the first part, we conducted experiments where a set of 50 individuals were shown approximately 150 synthetic graphs and 8 real-world networks and asked to provide their estimate of the strength of those networks. The process is explained in detail in section \ref{humansubject}.  The right-hand side of the strength metric defined in the previous section
($\sigma(G, W) = \sum_{i = 1}^n i \times w_i \times nc_i $)
is a linear equation with unknown variables $w_1, \ldots, w_n$. Once we have an estimate of the strength of a network from a participant, we assign it to the left-hand side of the metric. This provides us with an equation involving $n$ variables $w_, \ldots, w_n$.  The details of the process computing the values $w_i, \ldots, w_n$ are described in section \ref{eqnsolve}.

\subsubsection{Human Subject Estimate of Strength of Networks}
\label{humansubject}


To ensure that our metric aligns with human perception, we conducted a user study in which participants were presented with a variety of synthetic and real-world graphs mentioned above. The graphs were chosen to span a wide range of structural complexities: from highly fragmented graphs with many small components to more cohesive graphs with one or two large components.

Each participant estimated the perceived ``strength'' (or robustness, cohesion, connectedness, etc.) of the graph on a scale normalized to \([1, n]\), where \(n\) is the total number of nodes in each presented graph. For instance, a participant might assign a strength of \(1\) (very weak/fragmented) to a graph that appears to have many isolated nodes, and a strength close to \(n\) (very strong) to a fully connected or nearly fully connected graph.

Around 50 participants estimated the strength of each graph, and we aggregated their responses by taking the average. This average serves as the ground truth measure of ``perceived strength'' from a human perspective.

\begin{figure}[htbp]
    \centering
    \begin{subfigure}[t]{0.49\linewidth}
        \centering
        \includegraphics[width=\linewidth]{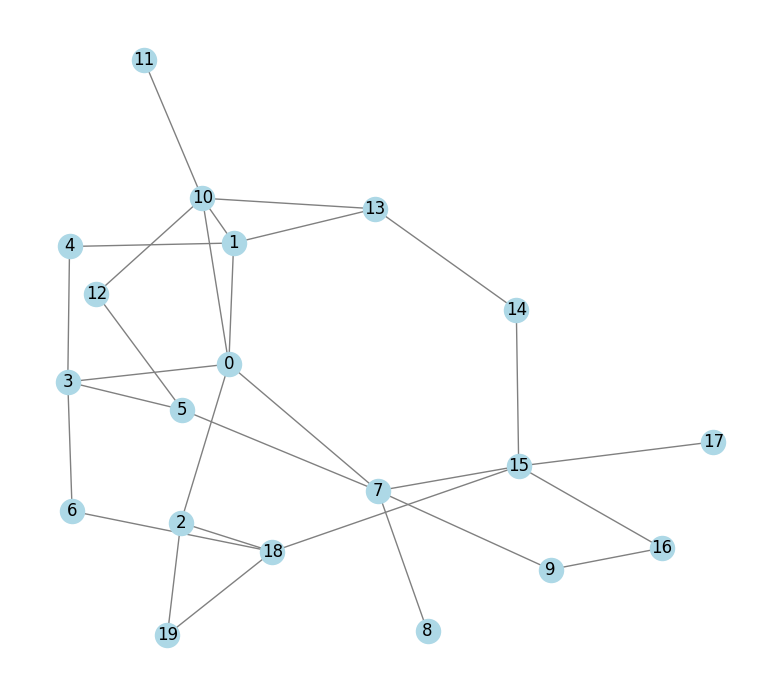}
        \caption{Sample Survey Graph (Connected).}
        \label{fig:sg1}
    \end{subfigure}
    \hfill
    \begin{subfigure}[t]{0.49\linewidth}
        \centering
        \includegraphics[width=\linewidth]{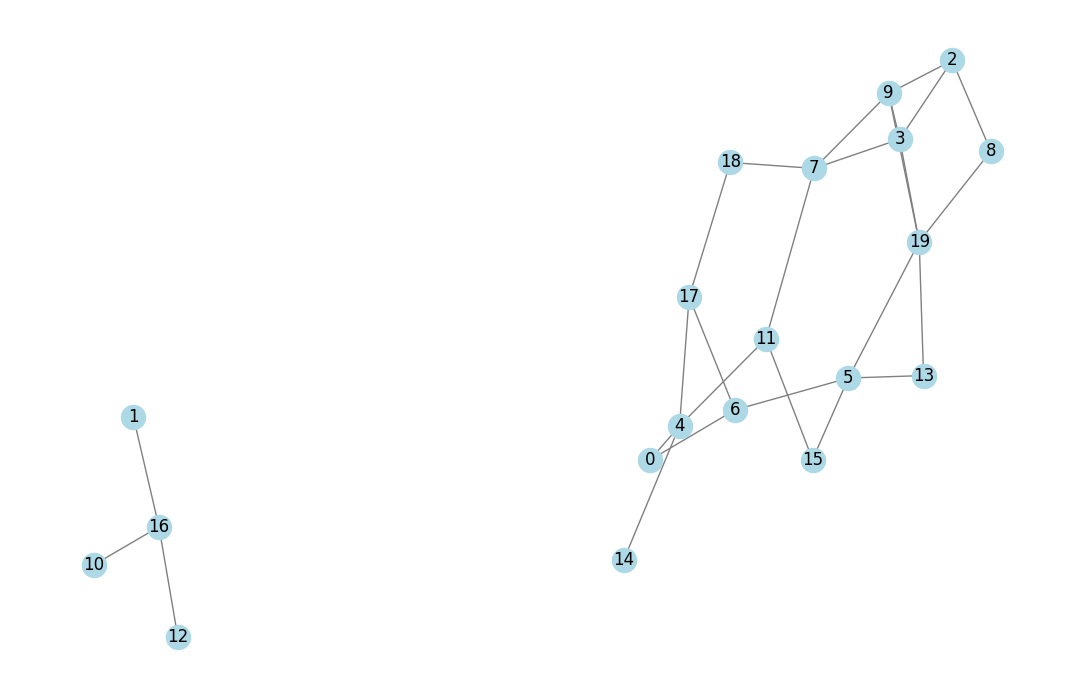}
        \caption{Sample Survey Graph (Disconnected).}
        \label{fig:sg2}
    \end{subfigure}
    \caption{Sample Graphs provided to Survey takers to get their perception of strength.}
    \label{fig:sg12}
    \vspace{-0.5cm}
\end{figure}

In Figure \ref{fig:sg12}, we can see two different graph networks out of 150 provided to the survey takers. They provided a value between 1 and 20 (total number of nodes) for both of these graphs as a measure of their respective strength.

\subsubsection{Formulating and Solving a System of Linear Equations}
\label{eqnsolve}

Given a set of \(m\) distinct graphs \(\{G_1, G_2, \ldots, G_m\}\) with human estimated values \(\{E_1, E_2, \ldots, E_m\}\), we express the estimate \(E_j\) for each graph \(G_j\) in terms of our metric:

\begin{equation}
E_j \;=\; \sum_{i=1}^n   i \times w_i \times nc_i,
\label{eq:system_equations}
\end{equation}

where \(nc_i\) is the number of connected components of size \(i\) in graph \(G_j\). Because each \(E_j\) is a known constant, we obtain \(m\) linear equations (one per graph). The unknowns are the weights \(\{w_1, w_2, \ldots, w_n\}\).

We solve this system of equations via least-squares regression, to find an optimal set of weights \(\hat{W}_1, \hat{W}_2, \ldots, \hat{W}_n\) that minimizes the discrepancy between the metric's computed strength and the average of human estimated values across all training graphs. In practice, having more diverse graphs (covering different component sizes) yields a more robust and stable solution.

To evaluate the model and derive meaningful insights from the graph structure, a system of linear equations was constructed and solved. The goal was to determine a set of best-fit weight values that align with the modeled behavior of the networks. The optimized weights for each variable ($\hat{W}_1$ to $\hat{W}_{30}$) are summarized in Table \ref{tab:variable_weights} below. 




\begin{table}[h!]
\centering
\begin{tabular}{|c c||c c||c c| }
\hline
\textbf{Variable} & \textbf{Weight} & \textbf{Variable} & \textbf{Weight} & \textbf{Variable} & \textbf{Weight} \\
\hline
$\hat{W}_1$  & 0.2221 & $\hat{W}_{11}$ & 0.6668 & $\hat{W}_{21}$ & 0.9193\\
$\hat{W}_2$  & 0.6607 & $\hat{W}_{12}$ & 0.7028 & $\hat{W}_{22}$ & 0.9847\\
$\hat{W}_3$  & 0.8747 & $\hat{W}_{13}$ & 0.8193 & $\hat{W}_{23}$ & 0.8122\\
$\hat{W}_4$  & 1.2271 & $\hat{W}_{14}$ & 0.7625 & $\hat{W}_{24}$ & 0.9321\\
$\hat{W}_5$  & 0.5538 & $\hat{W}_{15}$ & 0.9872 & $\hat{W}_{25}$ & 0.9485\\
$\hat{W}_6$  & 0.9078 & $\hat{W}_{16}$ & 0.7648 & $\hat{W}_{26}$ & 0.9868\\
$\hat{W}_7$  & 0.9445 & $\hat{W}_{17}$ & 1.0714 & $\hat{W}_{27}$ & 0.8559\\
$\hat{W}_8$  & 0.9517 & $\hat{W}_{18}$ & 0.6910 & $\hat{W}_{28}$ & 0.8390\\
$\hat{W}_9$  & 0.9737 & $\hat{W}_{19}$ & 0.9432 & $\hat{W}_{29}$ & 0.9867\\
$\hat{W}_{10}$ & 0.7178 & $\hat{W}_{20}$ & 0.8923 & $\hat{W}_{30}$ &  0.9093\\
\hline
\end{tabular}
\caption{Variable Weights}
\label{tab:variable_weights}
\vspace{-1cm}
\end{table}

These values serve as node weights in subsequent calculations involving custom strength metrics and graph fragmentation modeling. Their distribution also offers insights into the relative importance or influence of each node in the network under the defined model constraints.

\section{Integer Linear Program for Computation of $k$ Most Authoritative Nodes in the Network}

In this section, we provide an Integer Linear Programming (ILP) formulation for finding the $k$ most authoritative nodes in the network with a custom weight vector $W$.

Let $x_{ij} = 1$ if vertex $i$ is in component $j$. For each edge $\{i_1, i_2\} \in E$ you add a constraint that the vertices $i_1$ and $i_2$ have to be in the same component: $x_{i_1j} = x_{i_2j}$. Let $y_i$ be the vertex that gets disconnected. Constraints \ref{eq:edge_le} and \ref{eq:edge_ge} write the same thing as linearized inequalities, 
so that they are always satisfied if either $y_{i_1}$ or $y_{i_2}$ is 1. The first constraint in \ref{eq:assignC} ensures that each vertex is in exactly one component, and the second constraint assigns the value of $C_j$, which represents the size of each component. At max, $k$ vertices can be removed, which is ensured by the first constraint in \ref{eq:k_and_m}.  Constraints in \ref{eq:binary} ensure that $m_{jt}$, $x$ and $y$ are binary variables. $W$ is the weight vector. In Table \ref{tab:ilp_table}, the size is represented for each component. If we take the columnwise sum of that table, we get the number of components having each size from 0 to $n$. This table population and column-wise summation is handled by constraints in \ref{eq:k_and_m} and \ref{eq:C_and_S}. In the objective function \ref{eq:obj}, we minimize the strength of the graph and subtract the strength of the $k$ disconnected singleton components to simulate their removal from the graph.


\begin{table}[ht]
\centering
\begin{tabular}{|c|*{5}{c}|}
\hline
$m_{jt}$ & $t = 0$ & $t = 1$ & $\cdots$ & $t = n$ & \textbf{Summation} \\ \hline
$(j = 1)$ & $0$ & $1~(c_1 = 1)$ & $\cdots$ & $0$ & $\mathbf{1}$ \\ \hline
$(j = 2)$ & $1~(c_2 = 0)$ & $0$ & $\cdots$ & $0$ & $\mathbf{1}$ \\ \hline
$\vdots$ & $\vdots$ & $\vdots$ & $\cdots$ & $\vdots$ & $\vdots$ \\ \hline
$(j = n)$ & $0$ & $0$ & $\cdots$ & $1~(c_n = n)$ & $\mathbf{1}$ \\ \hline
\textbf{Summation} & $\mathbf{S_0}$ & $\mathbf{S_1}$ & $\cdots$ & $\mathbf{S_n}$ & \\ \hline
\end{tabular}
\caption{In this table, $t$ represents the size of components and $j$ represents the serial number of components.}
\label{tab:ilp_table}
\vspace{-1.5cm}
\end{table}

\begin{subequations}\label{eq:ILP}
\begin{alignat}{4}
\text{min}\quad
  & \left( \sum_{t=0}^{n} t \cdot S_t \cdot W_t \right)
     - \Bigl(\sum_{i=1}^{n} y_i\Bigr) \cdot W_1
     &&&&&&\label{eq:obj}\\[2pt]
\text{s.t.}\quad
  &x_{i_1j} \le x_{i_2j} + y_{i_1} + y_{i_2}
   && \forall\{i_1,i_2\}\!\in\!E,\ \forall j
   \label{eq:edge_le}\\
  &x_{i_1j} \ge x_{i_2j} - y_{i_1} - y_{i_2}
   && \forall\{i_1,i_2\}\!\in\!E,\ \forall j
   \label{eq:edge_ge}\\
  &\sum_{j=1}^{n} x_{ij} = 1             &&
   C_j = \sum_{i=1}^{n} x_{ij}
   && \forall i,\ \forall j
   \label{eq:assignC}\\
  &\sum_{i=1}^{n} y_i \le k              &&
   \sum_{t=0}^{n} m_{jt} = 1
   && \forall j
   \label{eq:k_and_m}\\
  &C_j = \sum_{t=0}^{n} t \cdot m_{jt}   &&
   S_t = \sum_{j=1}^{n} m_{jt}
   && \forall j,\ \forall t
   \label{eq:C_and_S}\\
  &m_{jt} \in \{0,1\}                    &&
   x_{ij},\,y_i \in \{0,1\}
   && \forall j,t,\ \forall i,j
   \label{eq:binary}
\end{alignat}
\end{subequations}





\section{Experimental Evaluation of the Proposed Metric}

\subsection{Results of Strength of Network Evaluation}
To validate the effectiveness of the proposed strength metric, a comparative analysis was conducted against three existing baselines: Cole 1, which is based on the number of connected components, Cole 2, which is based on the size of the largest connected component, and GFP, which is based on the attack score model. The Ground Truth values used in the evaluation were obtained by taking the average of the values estimated by the human subjects in the surveys. 

\subsubsection{Synthetic Graph Networks}
Here we use the synthetic graphs created using different NetworkX models as mentioned in Section \ref{sec:rwgraphs} that were not used in the weight vector calculations. The results are summarized in Figure \ref{fig:gt_vs_c1_c2} and \ref{fig:gt_vs_gfp_pm} below.


\begin{figure}[htbp]
    \centering
    \begin{subfigure}[t]{0.49\linewidth}
        \centering
        \includegraphics[width=\linewidth]{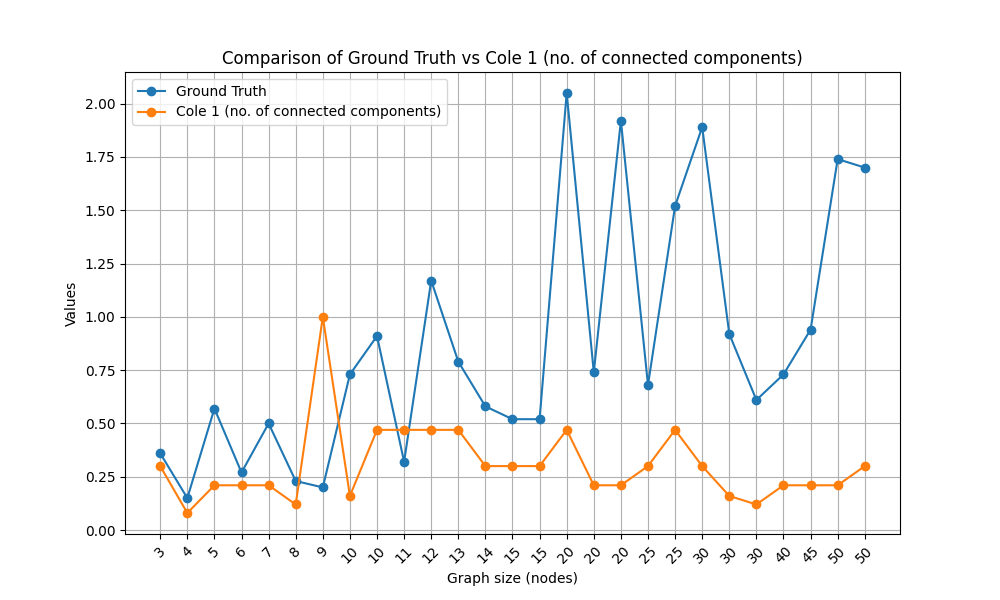}
        \caption{Comparison of Ground Truth and Cole 1 (Based on the Number of Connected Components).}
        \label{fig:gt_vs_c1}
    \end{subfigure}
    \hfill
    \begin{subfigure}[t]{0.49\linewidth}
        \centering
        \includegraphics[width=\linewidth]{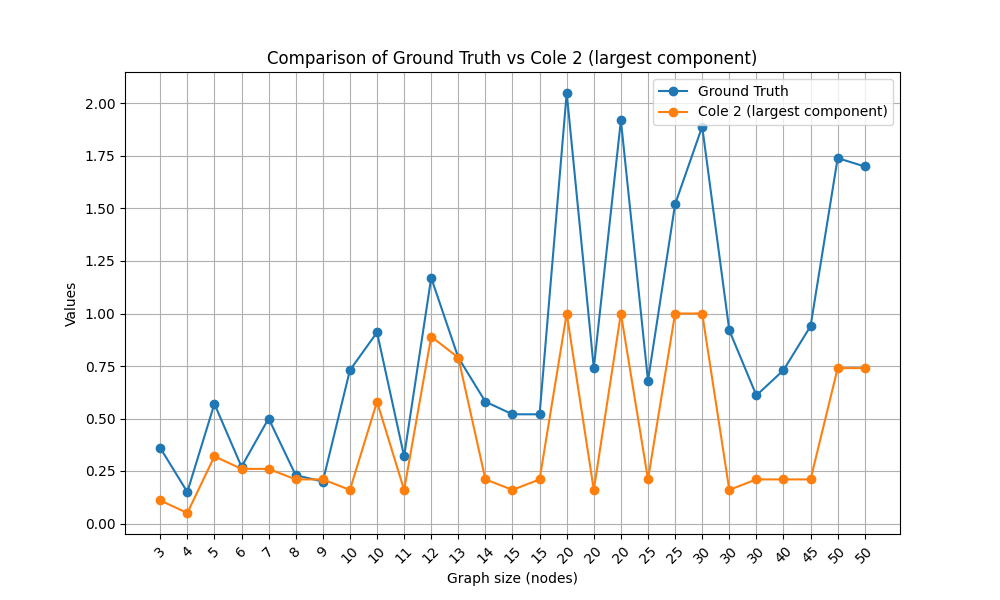}
        \caption{Comparison of Ground Truth and Cole 2 (Based on the Size of the Largest Component).}
        \label{fig:gt_vs_c2}
    \end{subfigure}
    \caption{Comparison between Ground Truth and Cole methods using two different metrics.}
    \label{fig:gt_vs_c1_c2}
\end{figure}






\begin{figure}[htbp]
    \centering
    \begin{subfigure}[t]{0.49\linewidth}
        \centering
        \includegraphics[width=\linewidth]{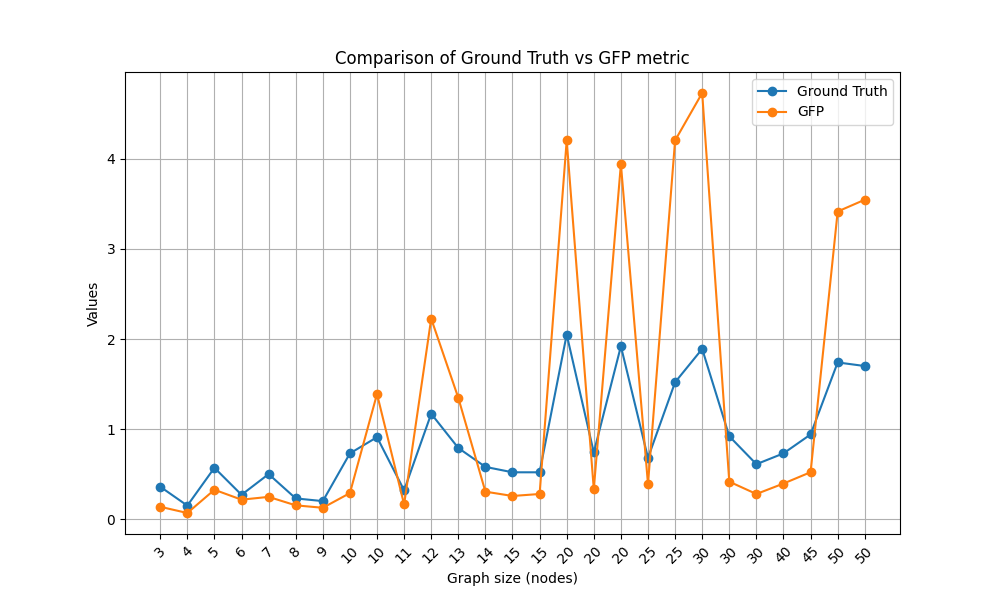}
        \caption{Comparison of Ground Truth and the GFP paper Metric.}
        \label{fig:gt_vs_gfp}
    \end{subfigure}
    \hfill
    \begin{subfigure}[t]{0.49\linewidth}
        \centering
        \includegraphics[width=\linewidth]{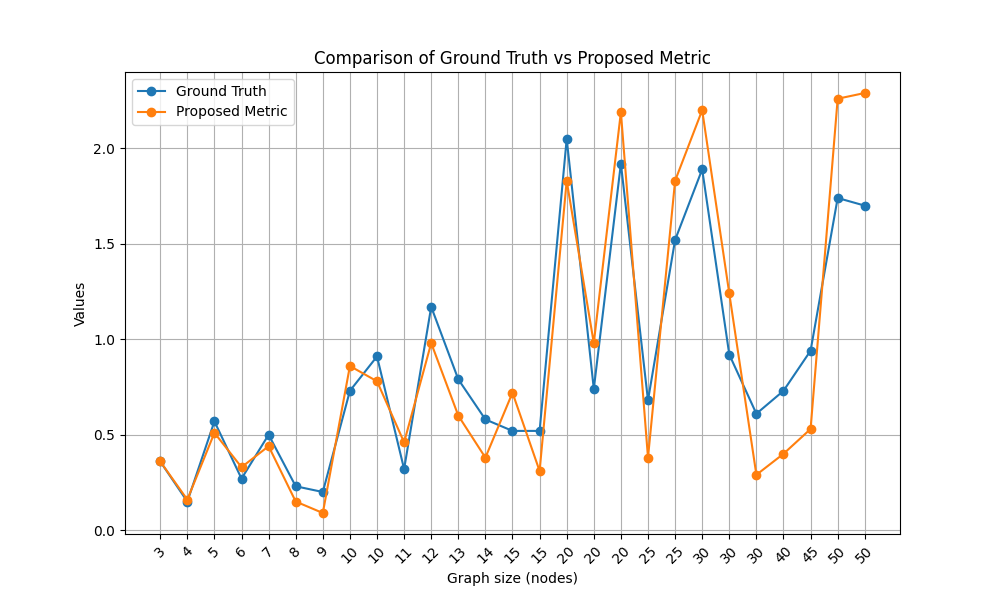}
        \caption{Comparison of Ground Truth and the Proposed Metric.}
        \label{fig:gt_vs_pm}
    \end{subfigure}
    \caption{Comparison of Ground Truth against GFP and Proposed metrics.}
    \label{fig:gt_vs_gfp_pm}
    \vspace{-0.5cm}
\end{figure}





Figure \ref{fig:gt_vs_c1_c2} presents a comparison of the Ground Truth values with the Cole 1 and Cole 2 methods on synthetic graphs generated using Erdős–Rényi and $G(n, m)$ models.

In Figure \ref{fig:gt_vs_c1}, we can see that the Ground Truth shows higher and more variable normalized strength values based on the number of nodes in the graph, particularly in medium-sized graphs (20–30 nodes). Cole 1 significantly underestimates this metric, producing a much flatter trend. This suggests that Cole 1 is insufficiently sensitive to the strength of larger graphs.
In Figure \ref{fig:gt_vs_c2}, Cole 2 significantly underrepresents strength values, especially for larger graphs. This indicates that Cole 2 may not scale effectively in random topologies where giant components emerge unpredictably.

Figure \ref{fig:gt_vs_gfp_pm} compares the Ground Truth against the GFP metric and the Proposed Metric.
In Figure \ref{fig:gt_vs_gfp}, the GFP metric deviates sharply from the Ground Truth, with frequent overestimations by a large margin. This suggests that the GFP metric is overly sensitive to strength in synthetic networks.
In contrast, in Figure \ref{fig:gt_vs_pm}, the Proposed Metric tracks the Ground Truth values closely across all node sizes. The alignment is particularly strong for denser graphs, demonstrating the adaptability and general consistency of the Proposed Metric in synthetic environments.

To quantitatively assess how well each metric aligns with the ground truth (obtained via human subject surveys), the Root Mean Squared Error (RMSE) was computed between the metrics' values and the ground truth values across all graph sizes. The results are summarized in Table \ref{tab:rmse}.


The RMSE values clearly indicate that the proposed metric exhibits the lowest error when compared with the ground truth, significantly outperforming both Cole's 1st and 2nd metrics as well as the GFP metric. The proposed metric achieves an RMSE of 0.2611, while Cole's 1st and 2nd metrics yield higher errors of 0.8070 and 0.5486, respectively while GFP having an error of 1.0763.

These findings further reinforce the qualitative observations from the visual comparisons. The proposed metric closely matches human perception, validating its potential as a reliable and interpretable strength measure for real-world networks.

\subsubsection{Real World  Networks}

\begin{figure}[htbp]
    \centering
    \begin{subfigure}[t]{0.49\linewidth}
        \centering
        \includegraphics[width=\linewidth]{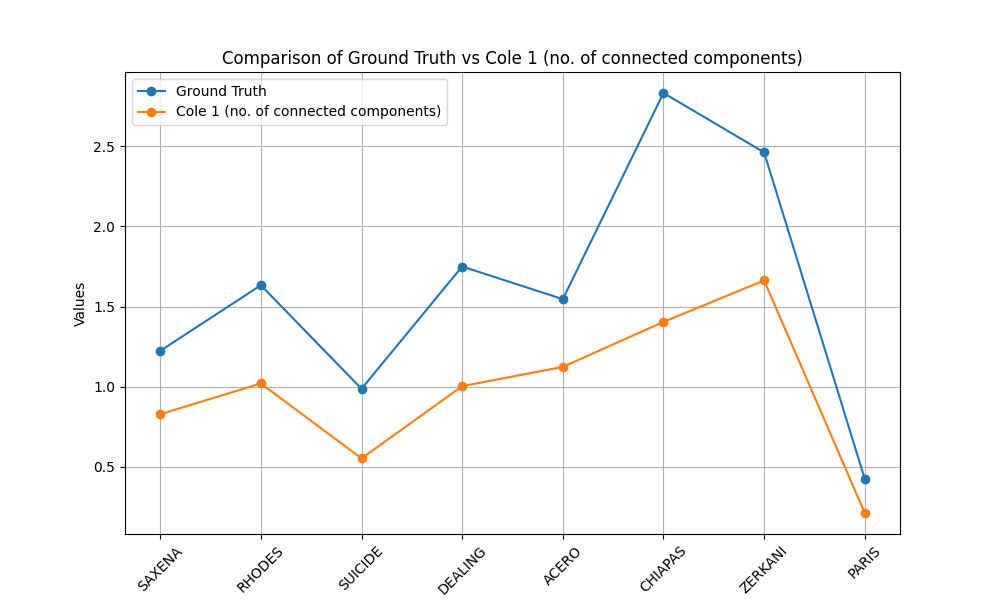}
        \caption{Comparison of Ground Truth and Cole 1 (Based on the Number of Connected Components).}
        \label{fig:rw_gt_vs_c1}
    \end{subfigure}
    \hfill
    \begin{subfigure}[t]{0.49\linewidth}
        \centering
        \includegraphics[width=\linewidth]{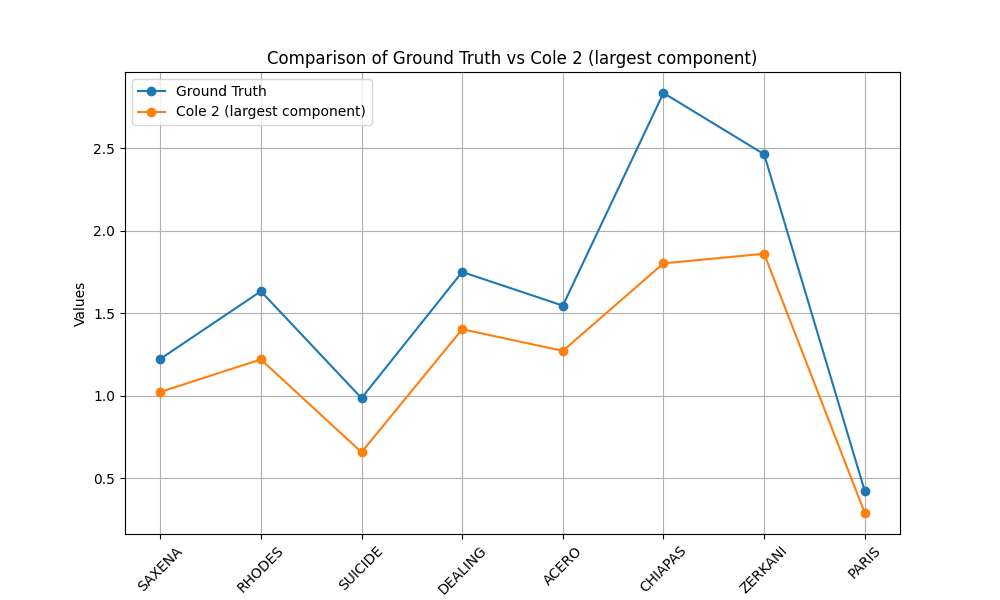}
        \caption{Comparison of Ground Truth and Cole 2 (Based on the Size of the Largest Component).}
        \label{fig:rw_gt_vs_c2}
    \end{subfigure}
    \caption{Comparison between Ground Truth and Cole's metrics.}
    \label{fig:rw_gt_vs_c1_c2}
\end{figure}

\begin{figure}[htbp]
    \centering
    \begin{subfigure}[t]{0.49\linewidth}
        \centering
        \includegraphics[width=\linewidth]{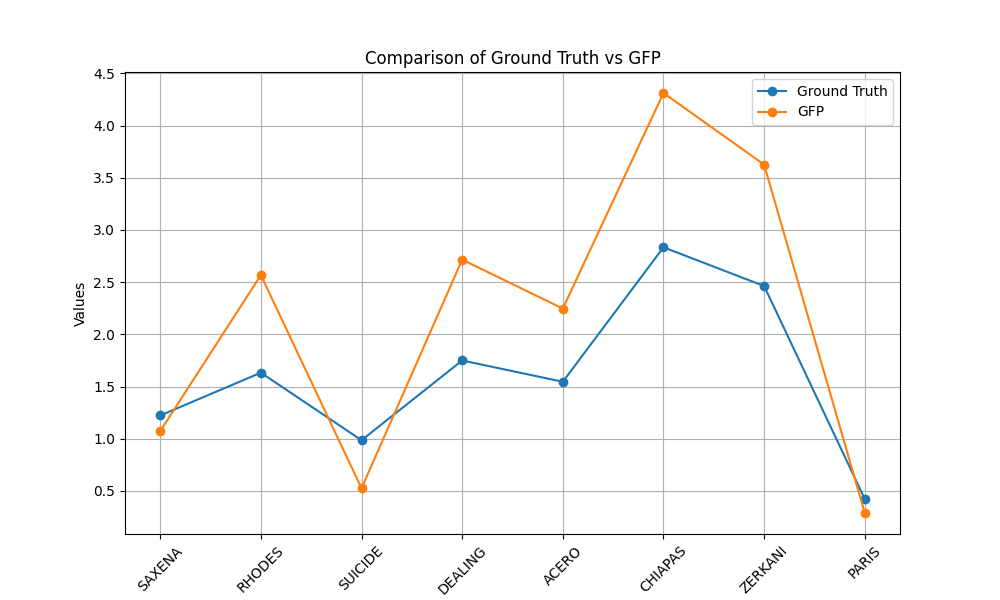}
        \caption{Comparison of Ground Truth and the GFP paper Metric.}
        \label{fig:rw_gt_vs_gfp}
    \end{subfigure}
    \hfill
    \begin{subfigure}[t]{0.49\linewidth}
        \centering
        \includegraphics[width=\linewidth]{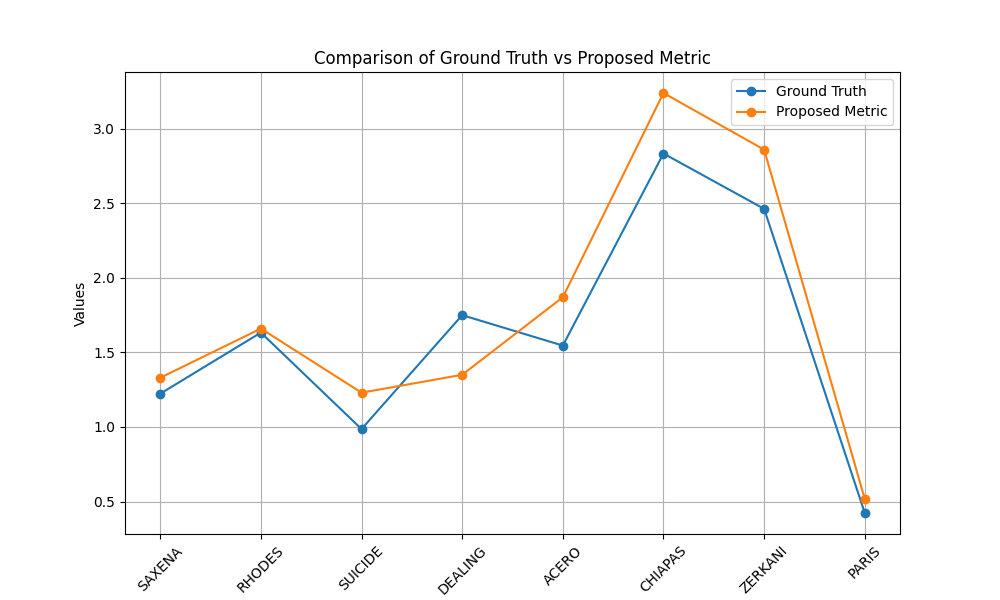}
        \caption{Comparison of Ground Truth and the Proposed Metric.}
        \label{fig:rw_gt_vs_pm}
    \end{subfigure}
    \caption{Comparison of Ground Truth against GFP and Proposed metrics.}
    \label{fig:rw_gt_vs_gfp_pm}
\end{figure}

Similar experiments were conducted on the real-world networks mentioned in Section \ref{sec:rwgraphs}. We present the results in Figure \ref{fig:rw_gt_vs_c1_c2} and \ref{fig:rw_gt_vs_gfp_pm}.

Figures \ref{fig:rw_gt_vs_c1_c2} and \ref{fig:rw_gt_vs_gfp_pm} illustrate the comparison results for real-world networks.
In Figure \ref{fig:rw_gt_vs_c1}, compared to the Ground Truth, Cole 1 generally underestimates the number of connected components, but the difference is less severe than in synthetic cases. This suggests that Cole 1 performs moderately better on structured, real-world topologies.
In Figure \ref{fig:rw_gt_vs_c2}, Cole 2 shows relatively close alignment with the Ground Truth, although it tends to slightly underestimate the strength of larger graphs. 
In Figure \ref{fig:rw_gt_vs_gfp}, the GFP metric again displays overestimation in most real-world cases, particularly in CHIAPAS and ZERKANI, indicating a lack of generalizability to practical graph data.
In Figure \ref{fig:rw_gt_vs_pm}, the Proposed Metric maintains close agreement with Ground Truth values across all networks. It effectively captures both the average case and outlier behavior, demonstrating its robustness across graph types.

This can also be confirmed by Table \ref{tab:rw_rmse}, where the Root Mean Squared Errors between ground truth and different metrics are presented.

To conclude, across both synthetic and real-world graph datasets, the Proposed Metric consistently outperformed existing methods in approximating the Ground Truth. While Cole-based methods lack sensitivity to larger graph strengths, and the GFP metric often exaggerates values, the Proposed Metric remained stable and accurate across a range of graph sizes and topologies. This suggests strong generalization capability and practical utility for real-world network analysis.

\begin{table}[htbp]
  \captionsetup[subtable]{justification=centering}
  \centering

  \begin{subtable}[t]{0.48\textwidth}
    \centering\footnotesize           
    \setlength{\tabcolsep}{4pt}       
    \begin{tabular}{|L{4.5cm}|r|}
      \hline
      \textbf{Metric} & \textbf{RMSE}\\\hline
      Ground Truth vs Proposed Metric & 0.2611\\\hline
      Ground Truth vs Cole's 1st Metric & 0.8070\\\hline
      Ground Truth vs Cole's 2nd Metric & 0.5486\\\hline
      Ground Truth vs GFP & 1.0763\\\hline
    \end{tabular}
    \caption{Synthetic graphs}
    \label{tab:rmse}
  \end{subtable}%
  \hfill
  \begin{subtable}[t]{0.48\textwidth}
    \centering\footnotesize
    \setlength{\tabcolsep}{4pt}
    \begin{tabular}{|L{4.5cm}|r|}
      \hline
      \textbf{Metric} & \textbf{RMSE}\\\hline
      Ground Truth vs Proposed Metric & 0.2890\\\hline
      Ground Truth vs Cole's 1st Metric & 0.7238\\\hline
      Ground Truth vs Cole's 2nd Metric & 0.4949\\\hline
      Ground Truth vs GFP & 0.8723\\\hline
    \end{tabular}
    \caption{Real‑world networks}
    \label{tab:rw_rmse}
  \end{subtable}

  \caption{Root Mean‑Squared Error (RMSE) comparison between ground truth and competing metrics on synthetic as well as real-world networks.}
  \label{tab:rmse_side}
\vspace{-1cm}
\end{table}

\subsection{Results of $k$ Most Authoritative Nodes Evaluation}

In this section, we conduct a comparative analysis of different metrics against the ground truth to evaluate their ability to find the $k$ most authoritative nodes in a network. For the sake of experimentation, we choose the value of $k$ to be 1 and 2. 

\subsubsection{Real World Networks}
We conducted experiments on the same real-world networks as in Section \ref{sec:rwgraphs} to identify the most authoritative nodes in those networks. Participants were asked to identify (i) the single most important/authoritative node in the network, and (ii) the two most important/authoritative nodes in the network. Accordingly, they identified the single most authoritative node and the two most authoritative nodes. This process of selecting one or two key nodes mirrors the decision-making approach of Law Enforcement Agencies (LEAs), who often focus on identifying the most critical individuals for surveillance or intervention based on perceived centrality or influence in the network.


Tables \ref{tab:single_node} and \ref{tab:two_nodes} summarize the survey results and their comparisons with the solutions obtained using our proposed network strength metric and three traditional metrics (Cole 1, Cole 2, and GFP). Multiple values in the Ground Truth columns reflect the varying percentage of responses from survey participants. The nodes/pairs of nodes are listed in order of percentage votes received in the survey responses.

We compare the closeness of our metric to the human perception with the traditional metrics using 3 different match parameters:

\begin{itemize}
\item \emph{Exact Match}: Measures the average accuracy of predictions exactly matching the top-ranked survey responses (the ground truth).
\item \emph{Rank Match}: Indicates the average rank at which predictions matched with the survey responses.
\item \emph{Percentage Match}: Represents the average percentage score, indicating how closely predictions align with participant responses.
\end{itemize}

\begin{table}[h!]
\centering
\begin{tabular}{|>{\centering\arraybackslash}p{1.3cm}|
                >{\centering\arraybackslash}p{2.2cm}|
                >{\centering\arraybackslash}p{1.5cm}|
                >{\centering\arraybackslash}p{1.0cm}|
                >{\centering\arraybackslash}p{1.0cm}|
                >{\centering\arraybackslash}p{1.0cm}|
                >{\centering\arraybackslash}p{2.4cm}|}
\hline
\textbf{Graph} & \textbf{Ground Truth} & \textbf{Our Metric} & \textbf{Cole 1} & \textbf{Cole 2} & \textbf{GFP} & \textbf{Metric} \\
\hline
SAXENA  & 4 2     & 4       & 2      & 2    & 2 &  \\ \hline
RHODES  & 14 16   & 16      & 16     & 16   & 16 &  \\ \hline
SUICIDE & 20      & 20      & 20     & 20   & 20 &  \\ \hline
DEALING & 14      & 14      & 14     & 14   & 14 &  \\ \hline
ACERO   & 23      & 23      & 23     & 23   & 23 &  \\ \hline
CHIAPAS &        21 13  & 21 & 13 & 13 & 13 & \\ \hline
ZERKANI &      20 8 5    & 8 & 5 & 8 & 8 & \\ \hline
PARIS &     6 8 4     & 6 & 6 & 4 & 4 & \\ \hline
        &         & \textbf{0.75}  & 0.5    & 0.375  & 0.375 & \textbf{Exact Match} \\ \hline
        &         & \textbf{1.25}  & 1.625    & 1.75  & 1.75 & \textbf{Rank Match} \\ \hline
        &         & \textbf{61.31} & 54.59  & 47.59 & 47.59 & \textbf{Percentage Match} \\
\hline
\end{tabular}
\caption{Comparison of Ground Truth vs. Our Metric and Traditional Methods for the single most authoritative node}
\label{tab:single_node}
\vspace{-0.8cm}
\end{table}

\begin{table}[h!]
\centering
\begin{tabular}{|>{\centering\arraybackslash}p{1.3cm}|
                >{\centering\arraybackslash}p{2.2cm}|
                >{\centering\arraybackslash}p{1.5cm}|
                >{\centering\arraybackslash}p{1.0cm}|
                >{\centering\arraybackslash}p{1.0cm}|
                >{\centering\arraybackslash}p{1.0cm}|
                >{\centering\arraybackslash}p{2.4cm}|}
\hline
\textbf{Graph} & \textbf{Ground Truth} & \textbf{Our Metric} & \textbf{Cole 1} & \textbf{Cole 2}  & \textbf{GFP} &  \textbf{Metric} \\
\hline
SAXENA  & {[2, 11] [2, 4]}                 & {[2, 11]}     & {[2, 11]}     & {[2, 11]}  &  {[2, 11]} & \\ \hline
RHODES  & {[16, 18] [14, 18] [14, 16] [5, 14] [5, 16] [1, 16]} & {[14, 16]}    & {[5, 16]}     & {[1, 16]}   & {[1, 16]} & \\ \hline
SUICIDE & {[11, 20] [1, 20] [15, 20] [4, 20] [2, 20]} & {[1, 20]}     & {[2, 20]}     & {[4, 20]}  &  {[4,20]} & \\ \hline
DEALING & {[14, 20] [14, 28] [14, 23]}    & {[14, 20]}    & {[14, 23]}    & {[14, 23]} &  {[14, 28]} & \\ \hline
ACERO   & {[5, 23] [1, 23]}                & {[1, 23]}     & {[1, 23]}     & {[1, 23]}   & {[1,23]} & \\ \hline
CHIAPAS &    [13, 21] [20, 21] [12, 21] [12, 20]      & {[20, 21]} & {[8, 20]} & {[8, 20]} & {[8, 20]} & \\ \hline
ZERKANI &       [20, 36] [8, 20] [20, 32] [20, 33]   & {[8, 20]} & {[5, 8]} & {[5, 8]} & {[5, 8]} & \\ \hline
PARIS &      [6, 8]     & {[6, 8]} & {[6, 8]} & {[6, 8]} & {[6, 8]} & \\ \hline

        &                                   & \textbf{0.375}  & 0.25  & 0.25 & 0.25 & \textbf{Exact Match} \\ \hline
        &                                   & \textbf{1.75}  & -           & -        & -  & \textbf{Rank Match}  \\ \hline
        &                                   & \textbf{38.15} & 23.95        & 24.91      & 25.85  & \textbf{Percentage Match} \\
\hline
\end{tabular}
\caption{Comparison of Ground Truth vs. Our Metric and Traditional Methods for the two most authoritative nodes}
\label{tab:two_nodes}
\vspace{-0.8cm}
\end{table}

As evidenced by Tables \ref{tab:single_node} and \ref{tab:two_nodes}, our proposed metric consistently outperforms the Cole 1, Cole 2, and GFP metrics across all measures—Exact Match, Rank Match, and Percentage Match. Particularly notable is the improvement in Exact Match and Percentage Match scores, highlighting our metric’s superior accuracy in identifying critical nodes according to human-subject consensus. For the Rank Match parameter in Table \ref{tab:two_nodes}, ``-" values represent that for some of the graphs, the result obtained by that metric didn't receive any vote in the respective survey, making it impossible to compute a valid rank.

These results validate the effectiveness of our metric in assessing critical nodes in covert network scenarios.

\section{Conclusion}



We proposed a novel metric for measuring network strength that combines structural properties with human perception, addressing limitations of traditional methods used in dismantling covert networks. By incorporating a tunable weight vector derived from the estimates provided by the human subjects, the metric aligns more closely with expert judgment in real-world scenarios.

Extensive evaluation on synthetic and real-world networks shows that our approach consistently outperforms existing metrics—Cole 1, Cole 2, and GFP. Notably, it achieves higher accuracy in identifying authoritative nodes, reflecting human consensus more effectively.

In future work, we plan to extend this approach through more diverse and structured experiments, including dynamic networks and application domains such as misinformation and epidemic response. These efforts aim to validate and generalize the utility of perception-aware strength modeling in broader contexts.

\bibliography{sn-article}

\end{document}